\documentstyle[prc,aps,epsfig]{revtex}

\draft 
\tighten

\def\be{\begin{equation}} 
\def\ee{\end{equation}} 
\def\bea{\begin{eqnarray}}
\def\eea{\end{eqnarray}} 
\def\nnb{\nonumber}

\begin{document}
\renewcommand{\thefootnote}{\fnsymbol{footnote}}
\setcounter{footnote}{1}

\title{
\hfill{\small {USC(NT)-Report-01-1}}\\
\hfill{\small {TRI-PP-01-02}}\\
\hfill{\small {SNUTP 00-037}}\\[0.2cm]
Polarized photons in radiative muon capture }

\author{Shung-ichi Ando$^a$\footnote{
E-mail address : sando@nuc003.psc.sc.edu},
Harold W. Fearing$^b$\footnote{
E-mail address : fearing@triumf.ca},
and Dong-Pil Min$^c$\footnote{
E-mail address : dpmin@snu.ac.kr}}

\address{$^a$ Department of Physics and  Astronomy,
University of South Carolina,
Columbia, SC 29208, USA} 
\address{$^b$ TRIUMF, 4004 Wesbrook Mall, Vancouver,
British Columbia, Canada V6T 2A3} 
\address{$^c$ School of Physics and Center for Theoretical Physics, Seoul 
National University, Seoul 151-742, Korea}

\date{April 24, 2001}

\maketitle

\begin{abstract}
We discuss the measurement of polarized photons arising
from radiative muon capture.
The spectrum of left circularly polarized photons or equivalently the
circular polarization of the photons emitted in radiative muon capture
on hydrogen is quite sensitive to the strength of the induced
pseudoscalar coupling constant $g_P$.
A measurement of either of these quantities, although very difficult,
might be sufficient to resolve the present puzzle resulting from the
disagreement between the theoretical prediction for $g_P$ and the
results of a recent experiment.
This sensitivity results from the absence of left-handed radiation
from the muon line and from the fact that the leading parts of the
radiation from the hadronic lines, as determined from the chiral power
counting rules of heavy-baryon chiral perturbation theory, all contain
pion poles.
\end{abstract}

\renewcommand{\thefootnote}{\arabic{footnote}}
\setcounter{footnote}{0}
\section{ Introduction}

The first measurement of radiative muon capture (RMC) on hydrogen,
\bea
\mu^-+p\to \nu_\mu + n + \gamma,
\eea
has been reported by a TRIUMF group \cite{rmc-experiment}, and the
value of the induced pseudoscalar constant $g_P$ was deduced to be
about 1.5 times larger than that predicted by the partially conserved
axial current (PCAC) or that obtained from one-loop order heavy-baryon
chiral perturbation theory (HBChPT) calculations
\cite{bernard-gp,fearing-omc}.
In Ref.~\cite{AM1} the photon spectrum from RMC on a proton was
obtained within the context of HBChPT up to next-to-next-to leading
order (NNLO), i.e., to one loop order. The results simply confirm a
next-to-leading order (NLO) HBChPT calculation \cite{meissner} and the
earlier theoretical predictions
\cite{opat,fearing-rmc,gmitro,beder1,beder2} based on a
phenomenological tree-level Feynman graph approach.  Furthermore, the
results of Ref.~\cite{AM1} indicated that the chiral series converges
rapidly, and thus suggest that the discrepancy between experiment and
theory observed for RMC on a proton cannot be explained by higher
order corrections within HBChPT.
Since then, many analyses have been reported, incorporating a variety
of new elements and suggestions, but all have essentially confirmed
earlier results and concluded that the existing discrepancy is still
unexplained
\cite{fearing98,bernard-fearing98,smejkal,bernard-omc-rmc,truhlik}.
\footnote{ A sea-gull term was introduced in the RMC amplitude in
Ref.~\cite{cheon}, which could reproduce the experimental data.
However, it was shown \cite{fearing-comment} that this term was not
gauge invariant and in addition that it was already present, together
with the additional pieces needed for gauge invariance, in the HBChPT
approach of Ref.~\cite{AM1} and in the standard Feynman graph method
of e.g. Ref.~\cite{fearing-rmc}.  }

As it appears that a NNLO calculation which includes all diagrams
through one-loop order converges sufficiently, the only possibilities
for significant improvement would seem to come from effects outside
the context of HBChPT, or perhaps from terms originating in the
Wess-Zumino Lagrangian.  These Wess-Zumino terms turn out to be
negligible however, as shown in Ref.~\cite{fearing98}.
Furthermore, all possible expressions in the amplitude which can be
composed of the characteristic operators involved in the reaction,
namely the polarization vectors of the photon and the lepton current,
the three-momenta of the outgoing photon and of the exchanged weak
vector boson, and the spin operator of the nucleon, emerge already in
the one-loop order.
Therefore, higher order contributions in the HBChPT perturbation
series will give corrections only to the coefficients of these
operator expressions and should be small, in view of the rapid
convergence of the chiral series in this reaction.
This led us to the conclusion that something other than the
ingredients of the hadronic vertices may in fact be the source of the
problem. For example, there may be difficulties in our understanding
of the atomic and molecular states of the muonic atom in hydrogen.
In particular the dependence of the photon spectrum on the initial
muonic atom states is non-negligible, so that it is important to try
to find a quantity which is less sensitive to the atomic and molecular
states but, at the same time, is sensitive to the pseudoscalar
constant.

Quite recently, some alternative scenarios for possible resolution of
the ``$g_P$ puzzle'' have been suggested by two groups.
In Ref.~\cite{AMK}, the photon spectrum corresponding to the
experiment of Ref.~\cite{rmc-experiment} was fitted by adjusting a
parameter $\xi$, with $(1-\xi)$ giving the fraction of spin $3/2$
ortho $p$-$\mu$-$p$ molecular state in liquid hydrogen.  A value
$\xi=0.8\sim 0.9$ was obtained, which is smaller than the theoretical
prediction $\xi=1$ \cite{halpern,bakalov} and would correspond to a 10
to 20 \% component of the spin 3/2 state
\footnote{The authors of Ref.~\cite{AMK} also considered ordinary, non
radiative muon capture (OMC) and originally found there the same value
of $\xi$ found for RMC. That result was obtained however using a
formula relating the liquid hydrogen and ortho molecular rates which
did not correspond to the experimental conditions of the OMC
experiment \cite{gorringe}.  Using an appropriate formula
\cite{omc-experiment,bardin} one finds that $\xi=1$ results in a value
which is in good agreement with the OMC data. However if one considers
the uncertainties in the data and in some of the parameters one finds
that values of $\xi$ as small as $\xi \sim 0.9$ are possible, which is
consistent, but only marginally so, with the result found for RMC.}.
In Ref.~\cite{bernard-omc-rmc}, on the other hand, the authors
speculate that the ``$g_P$ puzzle'' can be explained by accumulation
of small effects and variations of parameters, or perhaps by an
isospin breaking effect.

As we have observed, the present situation viewed from the context of
HBChPT can be summarized as follows. All symmetries of QCD are
respected order by order in this theory and the chiral expansion is
rapidly converging.
The rapid convergence is fortunate, since to improve the theory by
calculating higher orders would require including all of the many
possible diagrams of the chiral order under consideration and would
normally introduce a large number of new low energy constants which
would have to be constrained by experiments.
Furthermore, the HBChPT results agree fairly well with those obtained
from the standard diagram approach, so that all theoretical approaches
are reasonably consistent, and unable to explain the RMC data with the
predicted value of $g_P$.

It is probably important to remeasure the photon spectrum in RMC, or
to measure more precisely the rate for ordinary muon capture (OMC),
$\mu+p\to \nu +n$, as has been proposed \cite{psi}.  Alternatively,
one could consider performing a rather more sophisticated experiment
which would be sensitive to some different combination of the
ingredients of the problem.  In that vein, we want to propose here to
measure the polarization of the outgoing photon.

Measuring the photon polarization enables us to choose the most
important graphs which involve pion poles and therefore to enhance the
dependence of the result on the pseudoscalar coupling constant $g_P$.
In the usual transverse gauge\footnote{Note that individual diagrams
are not gauge invariant by themselves, so any comments about relative
sizes are gauge dependent. We will always assume the transverse gauge
for any such comparisons.}  by far the most important diagram for RMC
is the one where the photon is emitted from the leptonic current. The
pseudoscalar coupling constant is an important contributor to this
diagram, since $g_P$ is so much larger than $g_V$ or $g_A$, but its
importance is not enhanced by the pion pole because the momentum
transfer in this diagram is always spacelike.  Therefore, to
concentrate on the pseudoscalar constant, we would like to find the
channel where this diagram is blocked. The polarization experiment
blocks this channel.

The rationale is simple and transparent. Since the neutrino is
left-handed, the photon emitted from the leptonic current is
right-handed.  This was shown for V and A couplings in
Ref.~\cite{cutkosky} and generalized to include the induced couplings
as well in Ref.~\cite{fearing-theorem}.
A measurement of a left-handed photon filters out the photon from the
leptonic current, and is thus sensitive to radiation from the hadronic
current. The sensitivity to $g_P$ comes from the fact that some parts
of the hadronic current, and in particular some parts containing pion
poles, are of leading order by the power counting rules of HBChPT.

The photon circular polarization in RMC (to be defined explicitly
below) has been considered before in the context of a phenomenological
treatment of the weak nucleon current parameterized by form factors
\cite{fearing-theorem}. There it was shown that the circular
polarization (and also the photon asymmetry relative to the muon spin)
could be written as $1+{\cal O}(1/m_N^2)$ where $m_N$ is the nucleon
mass and where the coefficient of the ${\cal O}(1/m_N^2)$ term
involves the various coupling constants.  We will discuss below the
expansion scheme in powers of $1/m_N^2$ corresponding to this theorem
and its connection to the power counting scheme of HBChPT.

\section{Lepton matrix elements of RMC with
polarized photon}

The Feynman graphs contributing to RMC on a proton can be classified
into the two classes shown in Fig. 1: (a) the first corresponds to
those graphs where the muon radiates, and (b) the second to the graphs
where the hadron radiates.  The amplitude of the process can then be
written as the sum of two diagrams, 
\be 
M_{fi} = \frac{e G_F V_{ud}}{\sqrt{2}}\epsilon^*_\alpha \left[ 
{\cal M}^{\alpha\beta}J_\beta +{\cal J}_\beta M^{\alpha\beta} \right],
\label{eq;amplitude}
\ee 
where $e$ is the electric charge, $G_F$ is the Fermi constant,
$V_{ud}$ is a Kobayashi-Maskawa matrix element, and
$\epsilon^*_\alpha$ is the polarization vector of photon.  The hadron
matrix elements with three and four legs are denoted by $J_\beta$ and
$M_{\alpha\beta}$. Their properties have been studied in
Ref.~\cite{AM1}, and are briefly discussed in the next section.
\begin{figure}
\begin{center}
\epsfig{file=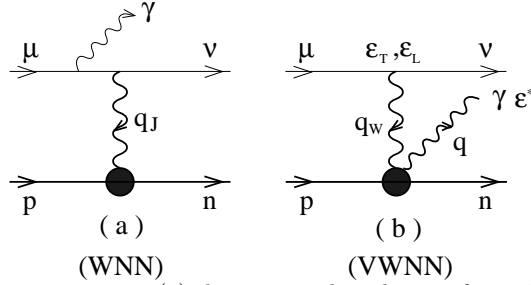,width=7cm} 
\caption{ Diagrams for radiative muon capture; (a) diagram with
radiation from the muon line.  The matrix element of the weak nucleon
current $J_\beta$ is matched with the lepton matrix element ${\cal
M}^{\alpha\beta}$.  (b) diagram with radiation from the hadronic
current whose matrix element $M_{\alpha\beta}$ is matched with the
lepton matrix element ${\cal J}^\alpha $.}
\label{fig;amplitude}
\end{center}
\end{figure}
The lepton matrix elements with three and four legs,
${\cal J}_\beta$ and ${\cal M}_{\alpha\beta}$ are
given by
\bea
{\cal J}_\beta &=& \bar{u}_\nu \gamma_\beta (1-\gamma_5)u_\mu,
\label{eq;jl}
\\
{\cal M}_{\alpha\beta} &=& \bar{u}_\nu\gamma_\beta(1-\gamma_5)
\frac{\gamma\cdot (\mu-q)+m_\mu}{2\mu\cdot q}\gamma_\alpha u_\mu,
\label{eq;ml}
\eea
where $\mu$ ($q$) is four momentum of muon (photon), $m_\mu$ is the
muon mass, and $u_\mu$ ($u_\nu$) is the Dirac spinor for the muon
(neutrino).

First, we study the lepton matrix elements involving a polarized
photon.  In the laboratory frame we assume that the $z$-axis of our
coordinate system coincides with the neutrino direction and the
$x$-$z$ plane includes the photon trajectory.  Thus we have 
\be
\hat{\nu}=(0,0,1),\ \ \ \hat{q}=({\rm sin}\,\theta,0,{\rm
cos}\,\theta), 
\ee 
where $\hat{\nu}$ ($\hat{q}$) is the unit vector of the neutrino
(photon) momentum and $\theta$ is the angle between neutrino and
photon, $\hat{\nu}\cdot \hat{q}={\rm cos}\,\theta$.  
In the transverse (Coulomb) gauge 
the polarization vectors of the photon are given by
\be 
\vec{\epsilon}^*_L=\frac{1}{\sqrt{2}}(-{\rm cos}\,\theta,-i,{\rm
sin}\,\theta), \ \ \ \vec{\epsilon}^*_R=\frac{1}{\sqrt{2}}({\rm
cos}\,\theta,-i,-{\rm sin}\,\theta),
\ee 
where subscripts $L$ and $R$ stand for the left- and right-handed
polarization state, respectively.  In this frame we can rewrite
Eqs. (\ref{eq;jl}) and (\ref{eq;ml}) in terms of components of four
vectors for each spin state,
\bea 
{\cal J}^\beta(+)
&\equiv& \varepsilon^\beta_{\rm T} = 2\sqrt{2m_\mu E_\nu} (0,-1,-i,0),
\label{eq;jp}
\\
{\cal J}^\beta(-)  &\equiv& \varepsilon^\beta_{\rm L}
= 2\sqrt{2m_\mu E_\nu} (1,0,0,1),
\label{eq;jm}
\\
{\cal M}^\beta(+,R) &=& 2\sqrt{\frac{E_\nu}{m_\mu}}
 (1+{\rm cos}\, \theta,{\rm sin}\,\theta,i\,{\rm sin}\,\theta,
  1+{\rm cos}\,\theta),
\label{eq;mpr}
\\
{\cal M}^\beta(-,R) &=& 2\sqrt{\frac{E_\nu}{m_\mu}}
 ({\rm sin}\, \theta,1-{\rm cos}\,\theta,i(1-\,{\rm cos}\,\theta),
  {\rm sin}\,\theta),
\label{eq;mmr}
\\
{\cal M}^\beta(\pm,L) &=& 0,
\label{eq;zero}
\eea
where ${\cal M}^\beta(\pm,h)\equiv \epsilon^*_{h,\alpha} {\cal
M}^{\alpha\beta}(\pm,h)$.  Signs ($\pm$) and $h$=($R$, $L$) in the
parenthesis of l.h.s. of the equations denote, respectively, up and
down muon spin state along the $z$-axis, and right- and left-handed
photon polarization state
\footnote{ We define the polarization vectors for the lepton current,
$\varepsilon_{\rm T}^\beta$ and $\varepsilon_{\rm L}^\beta$, as
depicted in Fig. \ref{fig;amplitude}, via Eqs. (\ref{eq;jp}) and
(\ref{eq;jm}).}.
Eqs.  (\ref{eq;mpr}), (\ref{eq;mmr}), and (\ref{eq;zero}) show
that the photons radiated from the muon line are totally
right-handed polarized \cite{cutkosky,fearing-theorem}.

If one measures the left-handed photons,
the amplitude of Eq. (\ref{eq;amplitude}) is reduced to
\be
M^{(L)}_{fi}=\frac{e\,G_F V_{ud}}{\sqrt{2}}
{\cal J}_\beta M^{\beta}(L) ,
\label{eq;right}
\ee
where $M^{\beta}(L)\equiv \epsilon^*_{L,\alpha}M^{\alpha\beta}$
is the part of $M^{\alpha\beta}$ producing only left-handed photons,
where the spin indices of proton and neutron are suppressed.

Therefore we can investigate the part of the hadron four-point matrix
element $M^\beta(L)$ which produces left-handed photons, without the
interference of the lepton radiating diagram containing the weak
nucleon current $J_\beta$, by measuring the left circularly polarized
photons.
The circular polarization $\beta$, which is defined by
\be
\beta \equiv \frac{N^R-N^L}{N^R+N^L}
\ee
where $N^R$ ($N^L$) is the spectrum of right-handed (left-handed)
photons\footnote{ The unpolarized spectrum $d\Gamma/dE_\gamma$ is
obtained by $d\Gamma/d E_\gamma=N^R+N^L$. }, has the property that
$\beta=1$ for the muon radiating diagram of  Fig.
\ref{fig;amplitude} (a) \cite{cutkosky,fearing-theorem}. Therefore,
for $\beta=1+\Delta\beta$, the deviation from one,
$\Delta\beta=-2N^L/(N^L+N^R)$, should come entirely from the
contribution of $M^{\beta}(L)$.

\section{Chiral counting rule and hadron matrix
elements of RMC}

HBChPT\cite{HBChPT} is a low energy effective field theory of QCD,
which has a systematic expansion scheme in terms of $Q/\Lambda_\chi$,
where $Q$ is a typical four-momentum scale characterizing the process
in question, $\Lambda_\chi$ is the chiral scale with
$\Lambda_\chi\simeq 4\pi f_\pi\sim m_N\simeq$ 1 GeV, and where $f_\pi$
is the pion decay constant. $Q$ must be small, typically of the order
of the pion mass $m_\pi$.
A typical scale $Q$ in muon capture (both OMC and RMC) is the muon
mass $m_\mu=105.7$ MeV, and hence $Q/\Lambda_\chi\simeq$ 0.1. One
therefore expects a rapid convergence of relevant chiral perturbation
series for muon capture and the explicit HBChPT calculations are
consistent with this expectation
\cite{bernard-gp,fearing-omc,AM1,meissner,bernard-omc-rmc,AMK}.

The effective Lagrangian is expanded as
\bea
{\cal L} = \sum {\cal L}_{\bar{\nu}}
 ={\cal L}_0 + {\cal L}_1 + {\cal L}_2 + \cdots,
\eea
where the subscript $\bar{\nu}$ denotes the order of terms,
$\bar{\nu}=d+n/2-2$, with $n$ the number of nucleon lines and $d$ the
number of derivatives or powers of $m_\pi$ involved in a vertex.
${\cal L}_0$, ${\cal L}_1$, and ${\cal L}_2$ are the leading order
(LO), next-to leading order (NLO), and next-to-next-to leading order
(NNLO) parts of the Lagrangian, respectively, and their explicit form
has been given in Ref.~\cite{AM1}. In passing, we should note that
the ${\cal L}_1$ includes the terms of ${\cal O}(1/m_N )$ which are
corrections to the leading order Lagrangian. In the NNLO Lagrangian we
have seven unknown constants, the so-called {\it low energy constants}
(LEC's), which are not determined by symmetry but must be fixed by
experiments.  Three of the seven LEC's appear in the three point
vertex functions of $J_\beta$, and they are fixed by the vector and
axial vector radius and the Goldberger-Treiman discrepancy
\cite{fearing-omc,AM1,bernard-omc-rmc,AMK}. One of the remaining four
constants is fixed via a rare pion decay \cite{donoghue}, and the
remaining three constants are estimated using the $\Delta$(1232) and
$\rho$ saturation method \cite{saturation}.~\footnote{ Recently, these
LEC's have been determined using data for radiative pion
capture\cite{fearing-pi-capture} and employing the Lagrangian of Ecker
and Moj\v{z}i\v{s}\cite{EMLagrangian}.} Therefore there are no
undetermined parameters in the calculation.

\begin{figure}
\begin{center}
\epsfig{file=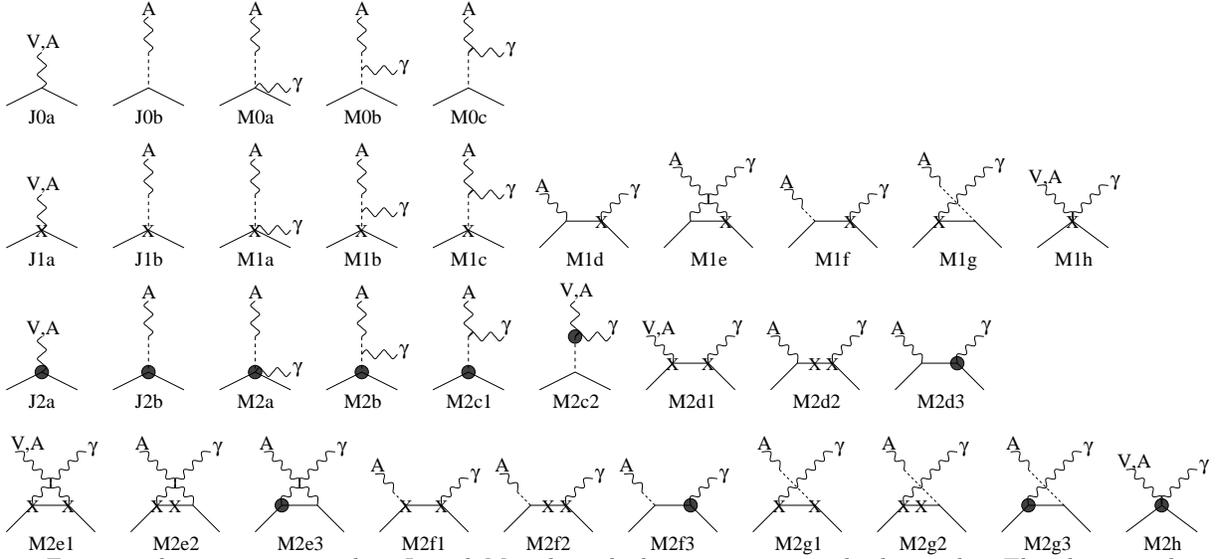,width=16cm}
\caption{Feynman diagrams contained in $J_\beta$ and $M_{\alpha\beta}$
through the next-to-next-to leading order.  The photon is denoted by
$\gamma$, the weak current is decomposed into $V,A$ parts, and a
dashed line denotes an exchanged pion.  Vertices without blobs are from
${\cal L}_0$, those with ``X'' are from ${\cal L}_1$ and those with
``$\bullet$'' are from ${\cal L}_2$ or the one-loop corrections.  The
five diagrams in the first line are diagrams of LO, which originate
from ${\cal L}_0$.  The next ten diagrams in the second line are those of
NLO, which contain one ``X'' of ${\cal L}_1$.  The following 19
diagrams in the third and forth lines are those of NNLO, which contain
two ``X''s of ${\cal L}_1$ or one ``$\bullet$'' of ${\cal L}_2$ or a
one-loop correction.  }
\label{fig;rmc}
\end{center}
\end{figure}

Let us look at the diagrams involving the hadron matrix elements
$J_\beta$ and $M_{\alpha\beta}$ in Fig. \ref{fig;rmc}.  (See the
caption of the figure for more details.)  The LO, NLO, and NNLO
diagrams are drawn in the first line, the second line, and the third
and fourth lines in Fig. \ref{fig;rmc}, respectively. Since, as noted
earlier \cite{AM1}, the series converges well, we expect those
diagrams in the first line to be the most important.
Both left- and right-handed photons are emitted from the hadron matrix
element $M_{\alpha\beta}$, and all the leading order diagrams of
$M_{\alpha\beta}$ (M0a, M0b, M0c) contain a pion pole.

Observe that two different momentum transfers appear in the pion poles
in the M0 diagrams.  For M0c and the lower pole of M0b, the momentum
transfer $q_J=\mu-\nu-q$ is relevant. $q_J^2$ is always spacelike, has
no significant $E_\gamma$ dependence, and is generally $\sim
-m_\mu^2$. On the other hand for the M0a diagram and the upper pole of
M0b the relevant momentum transfer is $q_W=\mu-\nu$. This depends on
$E_\gamma$ via $q_W^2\simeq 2m_\mu E_\gamma-m_\mu^2$ and becomes $\sim
+m_\mu^2$ near the upper end of the photon spectrum. Thus one is much
closer to the pion pole for these diagrams. This means that, other
factors being equal, these diagrams will be enhanced relative to those
involving $q_J$.

Now let us discuss the theorem of Ref.~\cite{fearing-theorem} and the
connection between the standard Feynman diagram approach to RMC and
the HBChPT approach described here.  In HBChPT the most important
diagram contributing to the hadronic pieces of Fig.  \ref{fig;rmc} is
the seagull diagram, M0a. This is just the standard Kroll-Ruderman
term, which however is not explicitly seen in the diagrams of the
relativistic phenomenological model (Fig. 1 of
Ref.~\cite{fearing-rmc}), since that model used a pseudoscalar
pion-nucleon coupling. Had pseudovector coupling been used it would
have appeared explicitly. It 
can however be directly identified as part of the diagram $M_b$
in Fig. 1 (b) of Ref.~\cite{fearing-rmc} where the photon radiates
from proton, the proton propagates, and interacts with the lepton
current, where the vertex of the weak nucleon current is described by
the weak form factors. The M0a diagram is included in the contribution 
from the negative energy propagation of the proton
in the $M_b$ diagram. (M0b and M0c can be also identified as parts of
(d) and (e) in Fig. 1 of Ref.~\cite{fearing-rmc}, respectively.)

In the phenomenological model the amplitude  $M_b$ can be expanded
in terms of $1/m_N$ as \bea M_b &=& \frac{1}{2m_N}\chi^\dagger_n
\left\{ g_V\left[\vec{\epsilon}^*\cdot\vec{\cal J}
+(1+\kappa_p){\cal
J}^0i\vec{\sigma}\cdot\vec{\epsilon}^*\times\hat{q}
-i\vec{\sigma}\cdot\vec{\epsilon}^* \times\vec{\cal J} \right]
\right. \nnb \\ && +g_A\left[{\cal
J}^0\vec{\sigma}\cdot\vec{\epsilon}^*
-(1+\kappa_p)(i\hat{q}\cdot\vec{\epsilon}^*\times\vec{\cal J}
+\vec{\sigma}\cdot\vec{\epsilon}^*\hat{q}\cdot\vec{\cal J}
-\vec{\sigma}\cdot\hat{q}\vec{\epsilon}^*\cdot\vec{\cal J})
\right] \nnb \\ &&\left. + g_P(q_W)\frac{q_W\cdot{\cal
J}}{m_\mu}\vec{\sigma}\cdot \vec{\epsilon}^*\right\}\chi_p+{\cal
O}(1/m_N^2), \label{eq;one_over_m} \eea where the nucleon weak
form factors are denoted by $g_V$ for vector, $g_A$ for axial
vector, and $g_P(q_W)$ for pseudoscalar form factors. $\kappa_p$
is the proton anomalous moment. We confirm the result of the
theorem \cite{fearing-theorem} that all the terms in Eq.
(\ref{eq;one_over_m}) are $1/m_N$ corrections. In this
approach the form factors are phenomenological parameters. The
$g_P$ dependent term is formally of order $1/m_N$, but the form
factor $g_P$ happens to be numerically large.

The connection to the HBChPT approach can be made via the
Goldberger-Treiman relation which tells us that the pseudoscalar form
factor has the structure due to pion propagation, i.e. a pion pole,
and is given explicitly by $g_P(q^2)=2 m_\mu m_N
g_A/(m_\pi^2-q^2)$. In HBChPT this expression, rather than $g_P$, will
appear in all the pion pole terms and the $m_N$ in the numerator will
cancel the $m_N$ appearing in the denominator, thus pushing this term
to one lower order in the expansion than it is in the expansion of the
phenomenological relativistic model\cite{fearing-theorem}.

We are now in a position to discuss what is known regarding the
polarization observables of the muon capture. As mentioned before, a
general theorem tells us that $\Delta\beta$ is formally ${\cal
O}(1/m_N^2)$\cite{fearing-theorem}. Using a phenomenological treatment
of the weak nucleon current parameterized by the form factors one can
show that hadron matrix elements are of order $J_\beta={\cal O}(1)$
and $M_{\alpha\beta}={\cal O}(1/m_N)$ in the $1/m_N$ expansion
\cite{fearing-theorem,fearing-ca}. Hence, $N^L\sim
|M^{\beta}(L)|^2={\cal O}(1/m_N^2)$ and the leading part of $N^R\sim
|J_\beta|^2={\cal O}(1)$ and thus $\Delta \beta={\cal O}(1/m_N^2)$ in
this model. However, $\Delta\beta$ is not particularly small, as also
noted in \cite{fearing-theorem}, because it contains a term
proportional to $g_P ^2$, and $g_P$ is large, as is explained in the
previous paragraph.

So to summarize, one can understand the connection between the theorem
derived by expansion of the relativistic phenomenological model in
Ref.~\cite{fearing-theorem} and the corresponding HBChPT expansion by
noting that there is a one to one correspondence between the $1$,
$1/m_N$, and $1/m_N^2$ terms in the expansion of the model and the LO,
NLO, and NNLO terms of HBChPT, except for the pion pole terms which
appear at one lower order in HBChPT because the $m_N$ in the numerator
of $g_P$ has been explicitly extracted.

\section{Numerical results}

\begin{figure}
\begin{center}
\epsfig{file=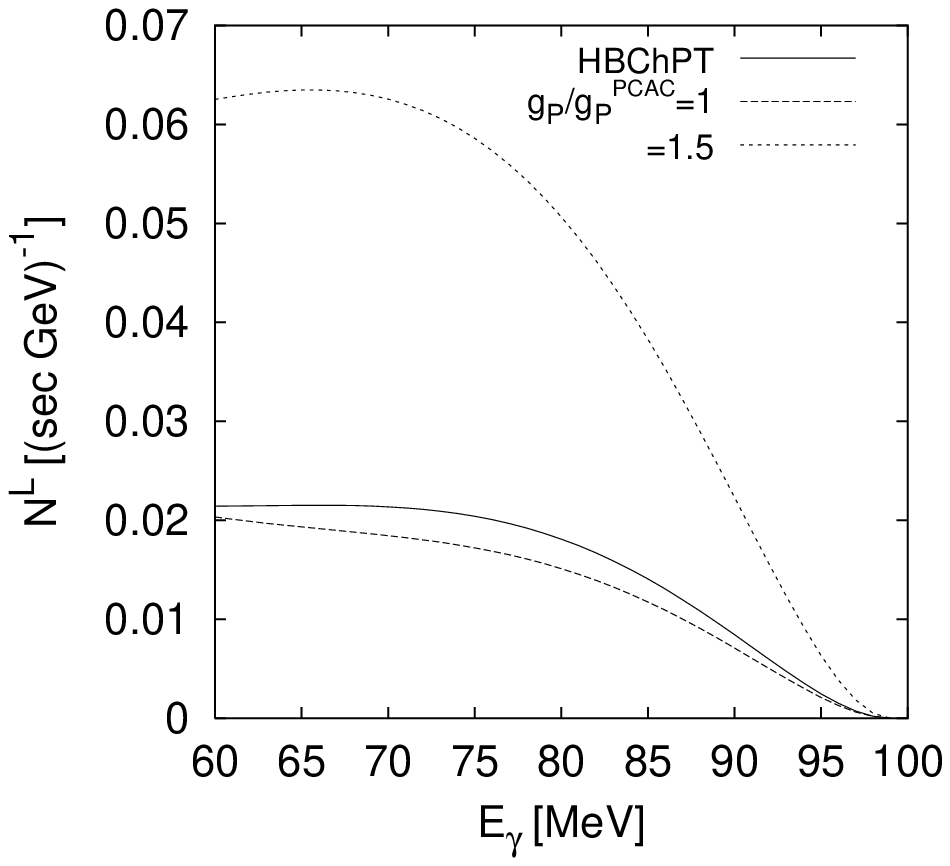, width=7.9cm}
\caption{The spectrum of left-handed photons
in the ``experimental state'' is plotted for the photon energy
$E_\gamma=$ 60 to 99 MeV.
The solid  line is the result of HBChPT up to the NNLO,
and the dashed and dotted lines are
results for  the relativistic model \protect\cite{fearing-rmc} with
two $g_P$ values, $g_P/g_P^{PCAC}=1$ and $1.5$, respectively.}
\label{fig;right-pol}
\vskip 10mm
\epsfig{file=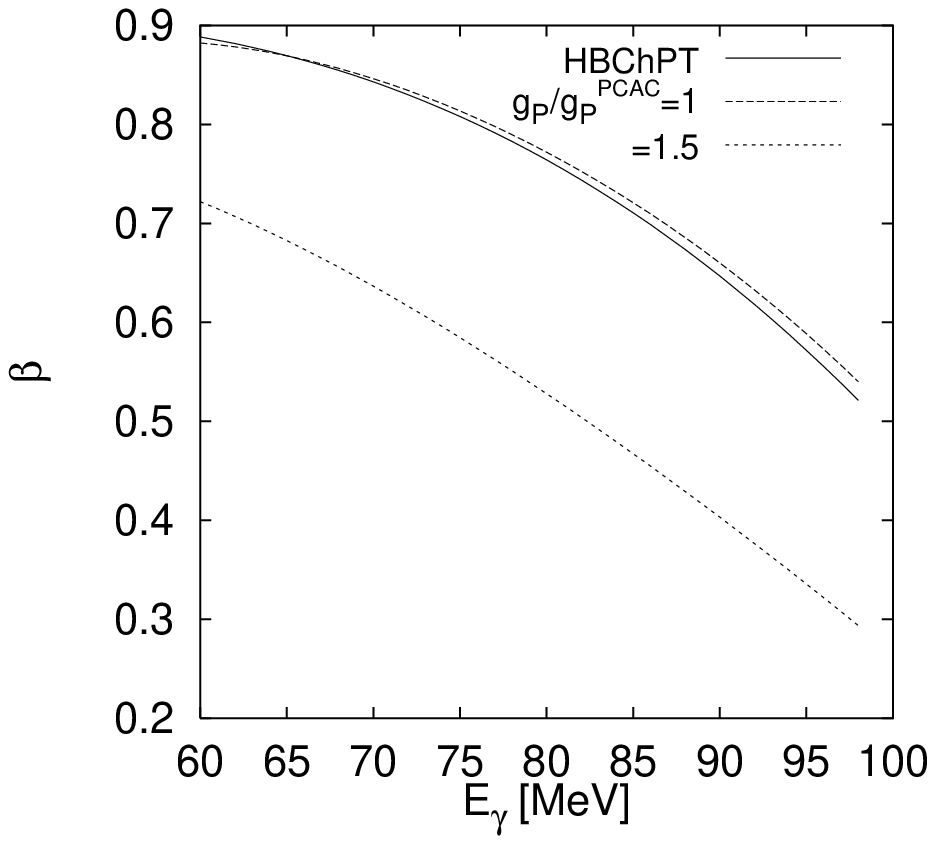, width=7.9cm}
\caption{
Circular polarization in the ``experimental state''.
See the caption of Fig. \ref{fig;right-pol}.}
\label{fig;circ-pol}
\end{center}
\end{figure}

\begin{figure}
\begin{center}
\epsfig{file=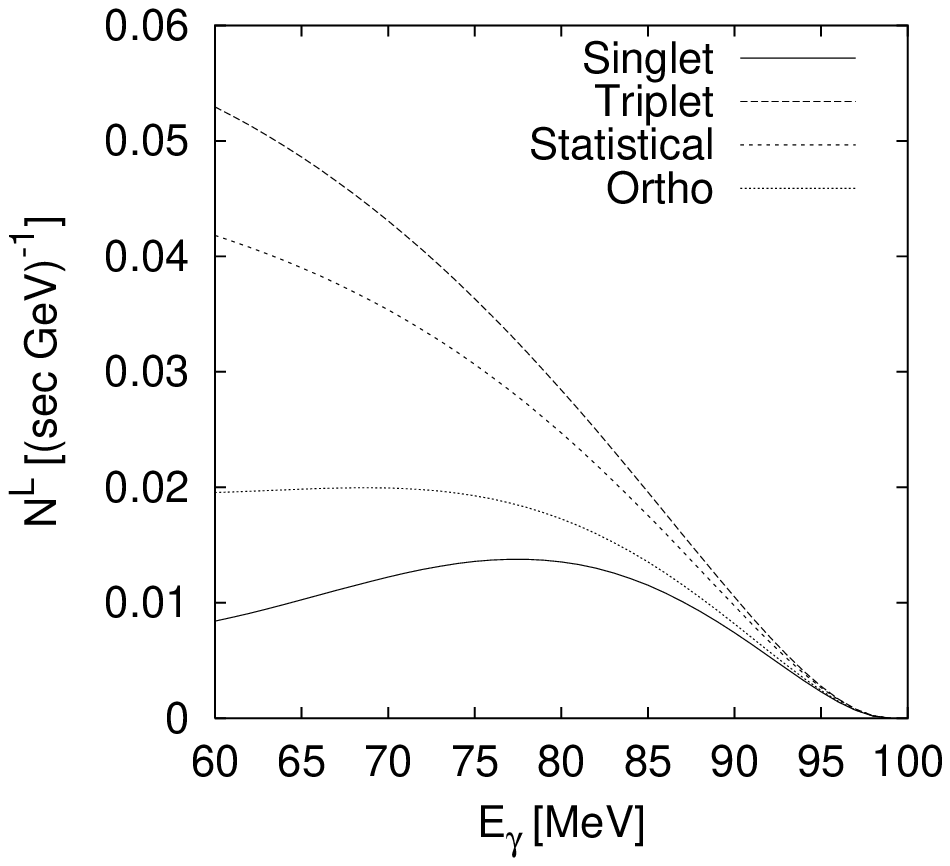,width=7.9cm}
\caption{The spectrum of left-handed photons is plotted for the photon
energy $E_\gamma= $ 60 to 99 MeV for each spin state.  The lines are
the results of HBChPT up to NNLO.  The solid, long-dashed,
short-dashed, and dotted lines correspond to singlet, triplet,
statistical, and ortho states, respectively.}
\label{fig;spectrum-left}
\vskip 10mm
\epsfig{file=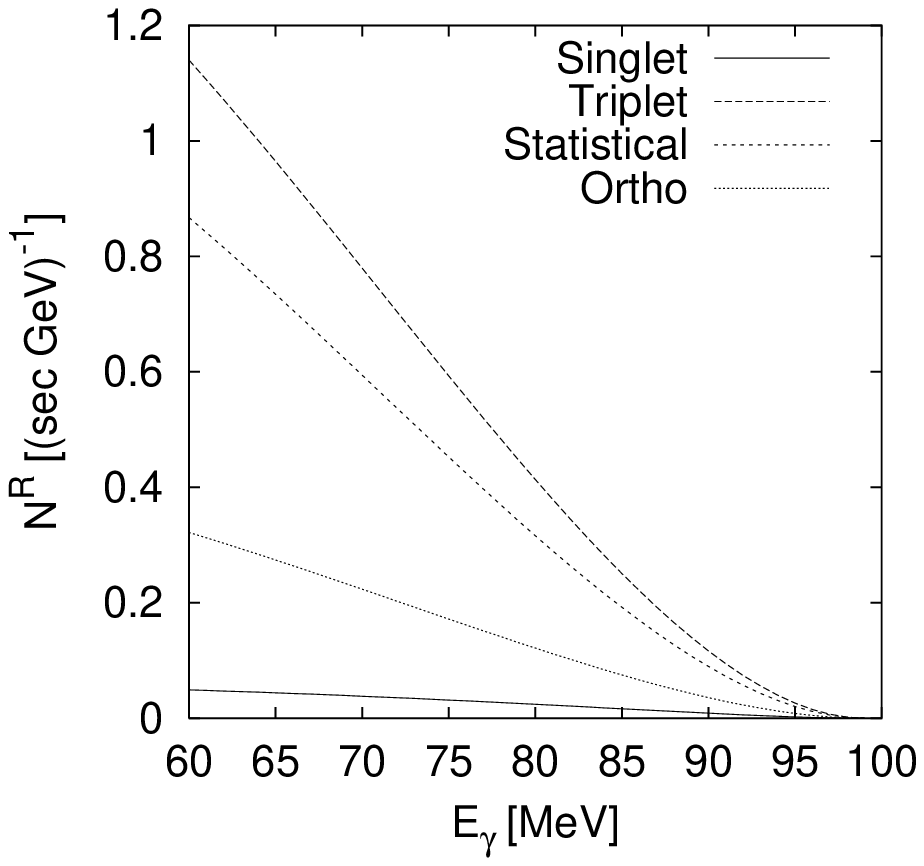,width=7.9cm}
\caption{ The spectrum of right-handed photons for each spin state.
See the caption of Fig. \ref{fig;spectrum-left}. }
\label{fig;spectrum-right}
\end{center}
\end{figure}

In Figs. \ref{fig;right-pol}, \ref{fig;circ-pol},
\ref{fig;spectrum-left} and \ref{fig;spectrum-right} we plot various
of our results for the spectrum and circular polarization of photons,
all calculated in HBChPT up to NNLO.  There are two major issues to
discuss. First, what is the sensitivity to $g_P$ of the spectrum of
left-handed photons and the circular polarization and, second, how
sensitive are these results to uncertainties in our knowledge of the
muon atomic states.

Let us first study the sensitivity of the polarization
observables to the value of $g_P$.
In Figs \ref{fig;right-pol} and \ref{fig;circ-pol}
we plot the spectrum of left-handed photons
and the photon circular polarization, respectively,
in the ``experimental state''
(6.1 \% atomic hyperfine singlet state,
85.4 \% ortho $p$-$\mu$-$p$ state,
and 8.5 \% para $p$-$\mu$-$p$ state)
reported in Ref.~\cite{rmc-experiment}
for the photon energy $E_\gamma=$ 60 MeV to 100 MeV.
We plot three lines which are obtained by using the HBChPT up to NNLO
and the relativistic phenomenological model \cite{fearing-rmc} with
two $g_P$ values, $g_P/g_P^{PCAC}=1$ and 1.5, where $g_P^{PCAC} \equiv
g_P(-0.88 m_\mu^2)$ is the Goldberger-Treiman prediction for $g_P$ at
the momentum transfer corresponding to OMC in hydrogen.

One finds that the results are quite sensitive to the value of $g_P$
as expected.
The results of HBChPT and the model with $g_P$=$g_P^{PCAC}$ are in
good agreement in the both figures which confirms that the same
basic ingredients are in both models and that the other higher
order corrections in HBChPT and terms not included in the
relativistic model are in fact small.
The case of $g_P/g_P^{PCAC}=1.5$ gives a photon spectrum larger by
about a factor of three than the case of $g_P/g_P^{PCAC}=1$.
Therefore our result shows the strong sensitivity of the polarized
photon spectrum to the different values of the pseudoscalar coupling
over the experimentally accessible photon energy region.  This is in
contrast to the unpolarized photon spectrum where the difference of
photon spectra with the two different values of $g_P$ is only of the
order of 30-40\% in the measurable region.
The circular polarization is also sensitive and differs for the two
values of $g_P$ by a more or less constant amount 0.2 over the whole
relevant region of photon energy.

Consider now the question of the sensitivity of the results to aspects
of the muon's atomic or molecular state.  The photon spectrum can
always be represented by a linear combination of the spectrum of
singlet and that of triplet state capture. The coefficient of each
state is determined by the particular target, liquid or gas, by the
amount of delay between the muon stop and the beginning of counting,
and by the formulas incorporating the various atomic and molecular
transition rates which describe the transitions from capture, through
singlet, ortho $p$-$\mu$-$p$ and para $p$-$\mu$-$p$ molecular
states. It is known that there are some ambiguities in the parameters
of these formulas, particularly with regard to the ortho-para
transition rate \cite{rmc-experiment} and to the possible inclusion of
a spin $3/2$ component in the ortho molecule \cite{AMK}.

In Figs. \ref{fig;spectrum-left} and \ref{fig;spectrum-right} we plot
our results for the spectra of left- and right-handed photons,
respectively, for each spin state. The solid, long-dashed,
short-dashed, and dotted lines correspond to singlet, triplet,
statistical, and ortho states, respectively.

>From these figures one can see immediately some general features. The
spectrum of right-handed photons, which is also essentially the
spectrum of unpolarized photons, is much larger than that for
left-handed photons. Specifically by comparing the two figures we find
that the rate for right-handed photons is about 2.5 times larger than
that for left-handed photons for the singlet state and 17.3 times
larger for the triplet state, when the spectra are integrated over the
photon energy $E_\gamma= $ 60 to 99 MeV. Under the experimental
conditions of the TRIUMF experiment \cite{rmc-experiment}, the ortho
molecular state is dominant, so that in these conditions one would
have about one-tenth as many left-handed photons as right-handed
ones. Presumably
this enhancement of right-handed photons is due to
the strong enhancement of the triplet state and to the fact that the
muon radiating diagram dominates, and, as was noted above produces
purely right-handed photons.

More specifically, with regard to the question of sensitivity to the
atomic and molecular states, we note that if the spectra of singlet
and triplet states were the same, the relative amounts would not
matter and there would be no sensitivity. From the figures we see
that, while this is not the case, the singlet is in fact much more
important, and closer to the triplet, for the left-handed photon case
than for the right-handed one. Numerically the ratio of the singlet to
triplet state spectra, when integrated over the photon energy, is 0.34
for left-handed photons and 0.05 for right-handed photons. This means
that the left-handed photon case will depend less strongly on the
relative amounts of singlet and triplet than the right-handed
case. But one should also take into account the result above that the
left-handed spectrum is much more sensitive to $g_P$ than the
right-handed (or unpolarized) spectrum. Thus one concludes that a
measurement of the spectrum of left-handed photons, or equivalently
the circular polarization of the photons, as we propose here, should
be significantly less sensitive to the atomic and molecular
ambiguities per unit of sensitivity to $g_P$ than is the right-handed
or unpolarized spectrum.

\section{Summary and discussion}

We have discussed RMC on the proton in the case when the measured
photon is polarized and have shown that the spectrum involving
left-handed photons and the photon circular polarization are quite
sensitive to the pseudoscalar coupling constant $g_P$.  They are
somewhat less sensitive than the unpolarized case to the atomic and
molecular spin state as well.
This is because the dominant diagram with radiation from the muon
vanishes when only left-handed photons are considered and because the
chiral counting rules of HBChPT select only the pion poles in the
leading order contribution from the other diagrams.
Thus these observables include the various ingredients of the problem
in a somewhat different way than does the unpolarized spectrum and so
their measurement may help resolve the current disagreement between
theory and experiment based just on the unpolarized spectrum.

The measurement of polarized photons in RMC on the proton is
technically extremely challenging.
The spectrum of left-handed photons is only one order of magnitude
smaller than that of the unpolarized photons.  However to measure the
polarization of the photon one needs an additional scattering through
an electromagnetic interaction or alternatively needs to measure the
angular distributions of the electron-positron pair produced when the
photon is stopped.
Hence to obtain the same order of precision as that of an unpolarized
RMC experiment, the polarization experiment must accumulate more
events, say by as much as four orders of magnitude, than the
unpolarized experiment. Such measurement is probably impossible with
current muon beams and techniques, but may become feasible with the
very intense muon beams which are now being discussed.

One should also note that there is an alternative quantity which could
be measured, namely the angular asymmetry of the photon relative to
the muon spin. By virtue of the general theorem of
Ref.~\cite{fearing-theorem} this quantity has generally the same
features and sensitivities as does $\beta$. It is much easier to
measure, since one does not need to rescatter the photon, and in fact
has been measured in nuclei \cite{virtue}. However in the case of the
proton, the muon loses almost all of its initial polarization as it is
captured into atomic orbit. Hence the suppression factor, due now to
the low residual polarization of the muon, may be just as large as for
the polarized photon observables we have considered here.

On the other hand, in nuclei the capture rate for RMC increases
proportional to $Z^4$, where $Z$ is the number of protons in the
nucleus.  This makes measurements of the unpolarized rate in nuclei
relatively easy \cite{armstrong,bergbusch}. So it may be feasible to
measure the polarized photon observables in RMC on heavy nuclei.
Indeed, the pion pole still gives the leading contribution and the
general features remain the same, although there are the not
insignificant complications in both calculations and interpretation
introduced by the nuclear structure.

\section{Acknowledgments}

We would like to thank M. Rho for his comments and helpful
discussions.  SA thanks T.-S. Park, F. Myhrer, and K. Kubodera for
comments and discussions.  DPM is very grateful to V. Vento for his
warm hospitality during his stay at University of Valencia. This work
is supported in part by Korea BK21 program,
by KOSEF 1999-2-111-005-5 and KRF Grant No. 2000-015-DP0072, 
by NSF Grant No. PHY-9900756 and INT-9730847, and
by the Natural Sciences and Engineering Research Council of Canada.

\end{document}